\documentstyle[a4,12pt,epsfig]{article}

\begin{document}
\begin{titlepage}

\begin{center}
{\Large \bf Budker Institute of Nuclear Physics}
\end{center}

\vspace{1cm}

\begin{flushright}
{\bf Budker INP 2001-80\\
November 21, 2001 }
\end{flushright}

\vspace{1.0cm}
\begin{center}{\Large \bf Experimental investigation of high-energy photon
splitting in atomic fields}\\
\end{center}
\vspace{1.0cm}

\begin{center}
{Sh.Zh.~Akhmadaliev, \fbox{G.Ya.~Kezerashvili}, S.G.~Klimenko,}\\
{R.N.~Lee,V.M.~Malyshev, A.L.~Maslennikov, A.M.~Milov,}\\
{A.I.~Milstein, N.Yu.~Muchnoi, A.I.~Naumenkov,
\fbox{V.S.~Panin},}\\
{ S.V.~Peleganchuk, G.E.~Pospelov, I.Ya.~Protopopov,}\\
{ L.V.~Romanov, A.G.~Shamov, D.N.~Shatilov,}\\
{ E.A.~Simonov, V.M.~Strakhovenko, Yu.A.~Tikhonov}\\

  Budker Institute of Nuclear Physics, Novosibirsk, 630090, Russia
\end{center}

\begin{abstract}
The new data analysis  of the experiment, where the photon
splitting in the atomic fields has been observed for the first
time, is presented. This experiment was performed at the tagged
photon beam of the ROKK-1M facility at the  VEPP-4M collider. In
the  energy region of 120-450~MeV , the statistics of $1.6\cdot
10^9$  photons incident on the BGO target was collected. About 400
candidates to the photon splitting events were reconstructed.
Within the attained experimental accuracy, the  experimental
results are consistent with the cross section calculated exactly
in an atomic field. The predictions obtained in the Born
approximation significantly differ from the experimental results.
\end{abstract}
\end{titlepage}

\section{Introduction}
Photon splitting is a process where the initial photon turns into
a virtual  electron-positron pair which  scatters   in the
electric field of an atom and then transforms  into two photons
sharing the initial photon energy $\omega_1$. This is an example
of  a self-action of an electromagnetic field, which results also
in such effects as coherent photon scattering (Delbr\"uck
scattering) and photon-photon scattering. The latter phenomenon
was never observed experimentally.

Delbr\"uck scattering was investigated in detail both
theoretically and experimentally \cite{PM75,MShu94,Akh98}. It
turned out that, for heavy atoms and high photon energy, its cross
section  calculated exactly in the parameter  $Z\alpha$ ($Z|e|$ is
the nucleus charge, $\alpha =\, e^2/4\pi\, =1/137$ is the
fine-structure constant) drastically differs from that obtained in
the lowest order in this parameter (Born approximation).

Recently, an essential progress in understanding of photon
splitting phenomenon was achieved. In papers \cite{LMS1,LMS2,LMS3}
various differential cross sections of high-energy photon
splitting have been calculated exactly in the parameter $Z\alpha$.
Similar to the case of Delbr\"uck scattering, the exact cross
section turns out to be noticeably smaller than that obtained in
the Born approximation. So, the detailed experimental
investigation of photon splitting provides a new sensitive test of
QED when the effect of higher-order terms of the perturbation
theory with respect to the external field is very important.

The observation of photon splitting is a difficult problem due to
a smallness of its cross section   as compared to those of other
processes initiated by   photons  in a target. The following
background processes are significant:  double Compton effect off
the atomic electrons ($\gamma e\to \gamma\gamma e$), and the
emission  of two hard photons from  $e^+e^-$ pair produced by the
initial photon. The relative importance of these processes depends
on the photon energy. For the energy  $\omega_1 \sim m$, where the
search of photon splitting was undertaken in two experiments
\cite{Adler66,Rob66}, only double Compton scattering determines
the background conditions. In these experiments,  the photons from
an intense radioactive source ($Zn^{65}$ with
$\omega_1~\simeq~1.1$~MeV in \cite{Adler66}, and $Co^{60}$ with
$\omega_1~\simeq~1.17,\, 1.33$~MeV in \cite{Rob66}) were used. The
combination of the coincidence and energy-summing detection
technique was applied. The number of events considered as
candidates for photon splitting exceeded the theoretical
expectations by the  factor of~300 in~\cite{Rob66}, and by the
factor of~6 in~\cite{Adler66}.

At high photon energy $\omega\gg m$ the emission  of hard photons
from  $e^+e^-$ pair becomes  most important as a background
process. In 1973 the experiment devoted to the study of Delbr\"uck
scattering of photons in energy region $1\div 7$~GeV was performed
\cite{DESY}. The bremsstrahlung non-tagged photon beam was used.
Some events were assigned by authors of~\cite{DESY} to the photon
splitting process. As shown in \cite{BKKF,dzhilk}, the theoretical
value for the number of photon splitting events under the
conditions of the experiment was two orders of magnitude smaller
than the experimental result. It was also argued that the events
observed could be explained by the production of $e^+e^-$ pair and
one hard photon.

The first successful observation of photon splitting  was
performed in 1995-96 using the tagged photon beam of the energy
$120\div450$~MeV at the VEPP-4M $e^+e^-$ collider~\cite{vepp} in
the Budker Institute of Nuclear Physics (Novosibirsk). Another
goal of this experiment was a study of Delbr\"uck
scattering~\cite{Akh98}. The total statistics collected
was~$1.6\cdot10^9$ incoming photons with~BGO (bismuth germanate)
target and $4\cdot10^8$ without target for background
measurements. The preliminary results were published in
\cite{Akh97}, \cite{ph98}. Here we present the new data analysis
for this experiment.

\section{Scheme of experiment}

\begin{figure}[htb]
\centering
\includegraphics*[width=0.5\textwidth]{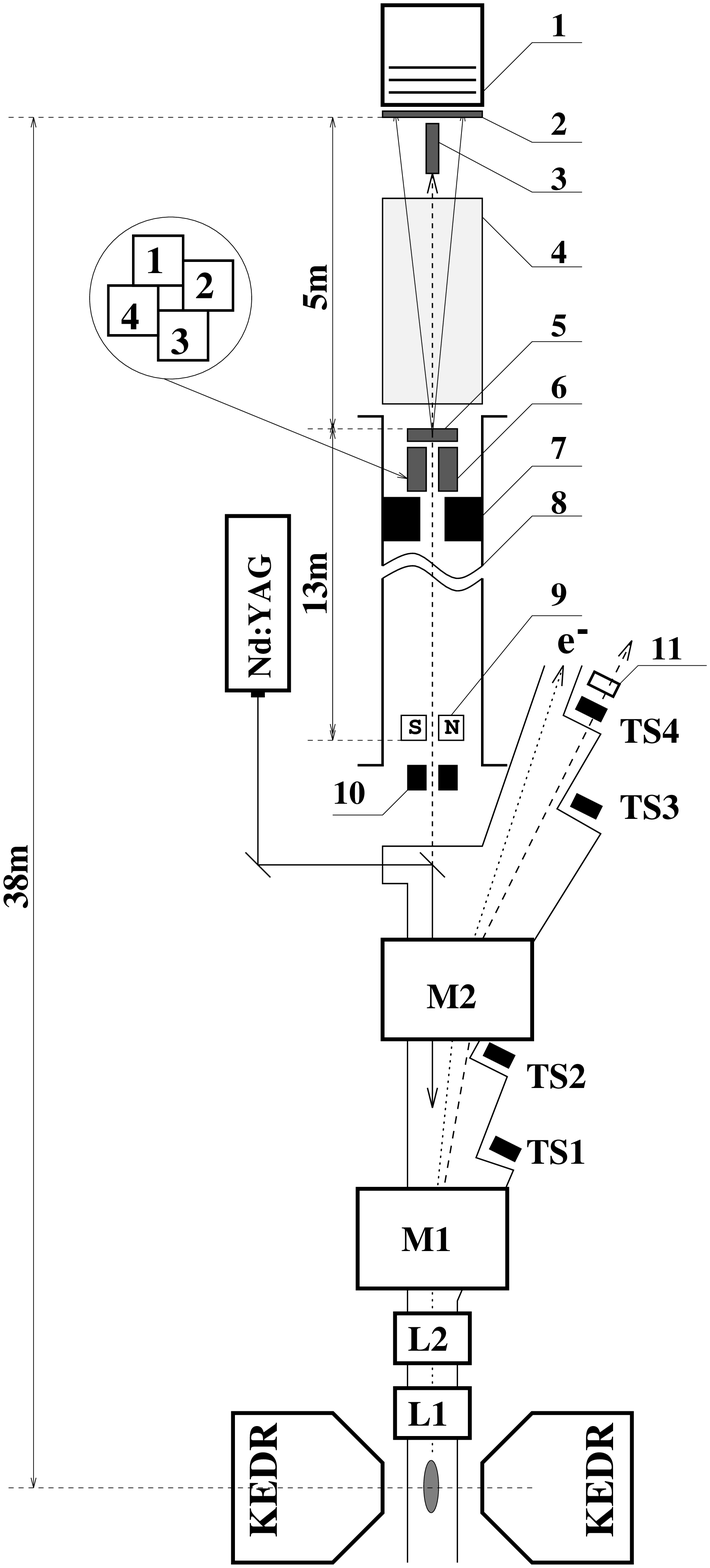}
\caption{Principal scheme of setup: LKr calorimeter (1);
scintillation veto-counter (2); beam  dump (BGO) (3); He-filled
tube (4m length) (4); target (BGO) (5); active collimator (BGO)
(6); lead absorber (7); guiding tube for the gamma-quanta beam
(8); cleaning magnet (9); passive lead collimator (10); TS
scintillattion counter (11); Nd:YAG is the laser; TS1-TS4 are
tagging system hodoscopes; M1 and M2 are bending magnets; L1 and
L2 are quadrupole lenses.} \label{Experiment:Fig:ps-setup}
\end{figure}

The experimental setup is shown in
Fig.~\ref{Experiment:Fig:ps-setup}. Some ideas of this setup were
suggested in~\cite{MW91}. The main features of the experimental
approach are:
\begin{itemize}
\item The use of high-quality tagged photon beam produced by
backward Compton scattering of laser light off high-energy
electron beam. Thereby, the energy of the initial photon is
accurately determined.
\item Strong suppression of the background processes by means of the detection
 of charged particles produced in the target and in other elements
 of the photon-beam line.
\item The detection of both final photons to discriminate the
photon splitting events from those with one final photon produced
in Compton or Delbr\"uck scattering.
\item The requirement  of the balance  between the sum of the  energies of the
final photons and the energy of the tagged initial photon. This
provides the  additional suppression of the events with charged
particles missed by the detection system.

\end{itemize}

At high energy of the initial photon $\omega_1 \gg m$, the photon
splitting cross section is peaked at small angles between momenta
of all photons ($\sim m / \omega_1$). Therefore, a good
collimation of the primary photon beam is required. The ROKK-1M
facility~\cite{ROKK} is used as the intense source of the tagged
$\gamma$-quanta. The electron energy loss in the process of
Compton scattering of laser light is measured by the tagging
system (TS)~\cite{TS} of the KEDR detector~\cite{kedr}. The TS
consists of the focussing spectrometer formed by accelerator
quadrupole lenses, bending magnets, and 4 hodoscopes  of the drift
tubes. High-energy photons move in a narrow cone around the
electron beam direction. The angular spread is of the order
$1/\gamma$, where $\gamma=E_{beam}/m$ is the relativistic factor
of the electron beam. The photon energy spectrum has a sharp edge
at
\begin{equation}
   \omega_{th}=\frac{4\gamma^2\omega_{laser}}{1+4\gamma\omega_{laser}/m}\,,
\end{equation}
that allows one to perform the absolute calibration of the tagging
system in a wide energy range. In experiment the laser photon
energy was $\omega_{laser}=1.165$~eV, the electron beam energy
$E_{beam}=5.25$~GeV, and $\omega_{th}=450$~MeV. The photon energy
resolution provided by the tagging system depends on the photon
energy and on the position of scattered electron in TS hodoscope:
it was 0.8~\% at $\omega_1=450$~MeV (at the center of the
hodoscope) and $\sim5$~\% at $\omega_1=120$~MeV (at the edge of
the hodoscope).
     The collimation of the photon beam is provided by two collimators
spaced at 13.5~m. The last collimator, intended to strip off the
beam halo produced on the first one, is made of four BGO (bismuth
germanate) crystals as shown in the separate view in
Fig.~\ref{Experiment:Fig:ps-setup}. After passing through the
collimation system, the photon beam  hits $1X_0$ thick BGO crystal
target. In order to separate the photons passed the target without
interaction from those scattered in the target, certain angular
region around the photon beam direction ($\theta \le 2.4$~mrad)
was enclosed by the dump. It is made of 13~$X_0$ thick BGO crystal
installed in front of the photon detector. The only photons to be
detected are those scattered outside the dump shadow. All active
elements used in the beam line (collimators, target, dump,
scintillating veto counter) set a veto signal in the trigger and
their signals are used in the analysis for background suppression.
The information from the target and beam dump is also used for
measurement of the incoming photon flux. The liquid-krypton
ionization calorimeter is used for the detection of the final
photons.  Its three-layer double-sided electrode structure enables
one to get both (X and Y) coordinates for detected photons. The
energy resolution of the calorimeter is~$2\% /
\sqrt{\omega(\mbox{GeV})}$. The liquid-krypton calorimeter is
described in details in \cite{prot,lkrpos}.

\begin{figure}[htb]
\centering
\includegraphics*[width=0.9\textwidth]{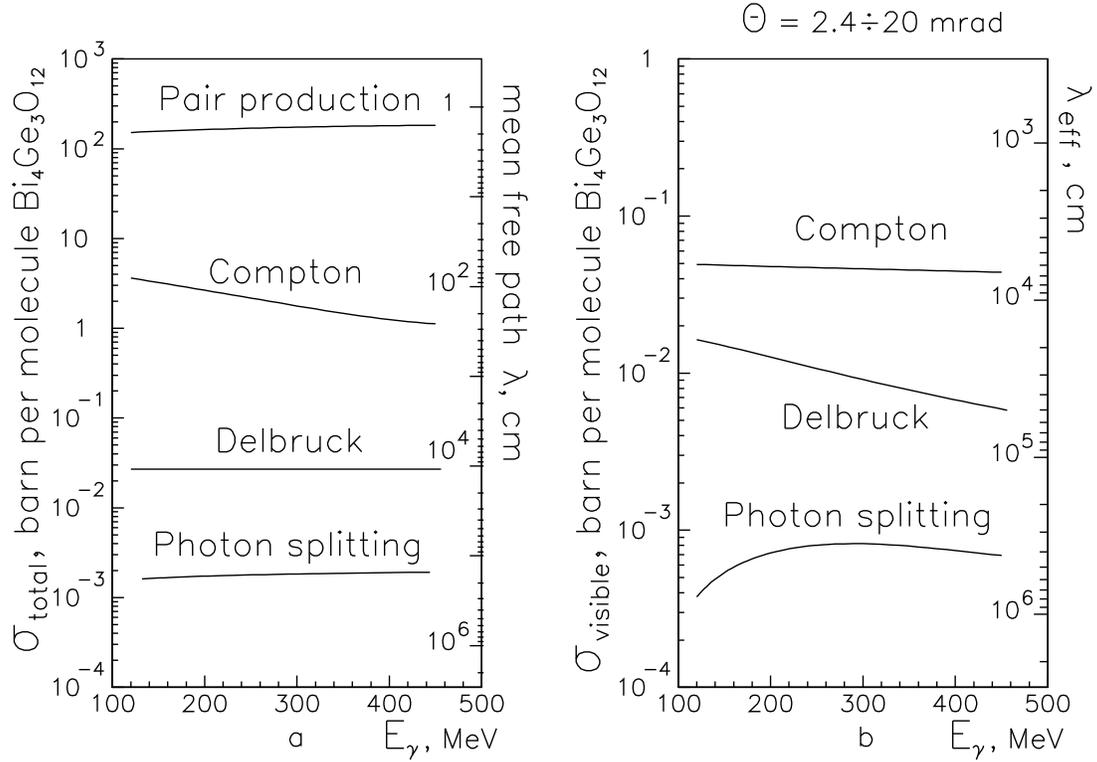}
\caption{The calculated energy dependence of the total (a) and the
visible (b) cross sections of various processes initiated by
photons in BGO~target (in units of barn per one molecule of
$Bi_{4}Ge_{3}O_{12}$).} \label{Experiment:Fig:crsec}
\end{figure}

In the experiment, the detected final photons had the polar angles
in the region $2.4~\mbox{mrad}\le \theta \le 20~\mbox{mrad}$. The
corresponding cross section is called  "visible".
Fig.~\ref{Experiment:Fig:crsec} shows the calculated energy
dependence of the total ($a$) and visible ($b$) cross sections for
various processes initiated by photons in BGO~target: $e^+e^-$
pair production, Compton scattering on atomic electrons,
Delbr\"uck scattering, and photon splitting. The calculation of
the photon splitting cross section was performed using the results
obtained in \cite{LMS1,LMS2,LMS3}.

\section{Results}

In the event selection procedure the following constraints   were
applied:
\begin{itemize}
\item The absence of the signal caused by charge particles in all active
elements of the photon-beam line.
\item The  balance of  the tagged initial
photon energy and the energy measured in the calorimeter within
$3\sigma$ of its energy resolution.
\item The existence of two separate tracks at least for one (X or Y) coordinate
 in the calorimeter strip structure.
\end{itemize}

The fulfillment of the latter  requirement strongly suppresses the
contribution of the processes with one photon in the final state
which could imitate two photon events in the calorimeter.

\begin{figure}[htb]
\centering
\includegraphics*[width=0.9\textwidth]{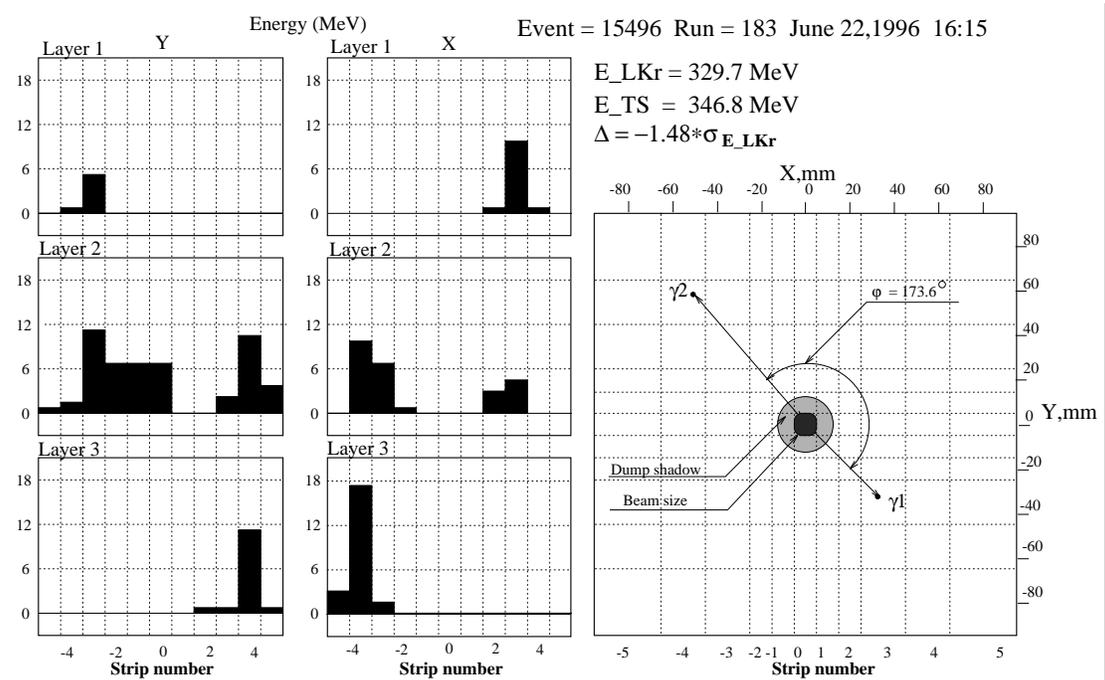}
\caption{Energy profile in the calorimeter strip structure (left)
and reconstructed kinematics (right) for a typical candidate to
the photon splitting event.} \label{Experiment:Fig:event}
\end{figure}

The typical event which meets selection criteria is shown in
Fig.~\ref{Experiment:Fig:event}. In this example two tracks are
seen in both X and Y directions. The conversion of the first
photon  occurs in the Layer 1 while the second photon converts in
the Layer 2.

\begin{figure}[h]
\centering
\includegraphics*[width=0.5\textwidth]{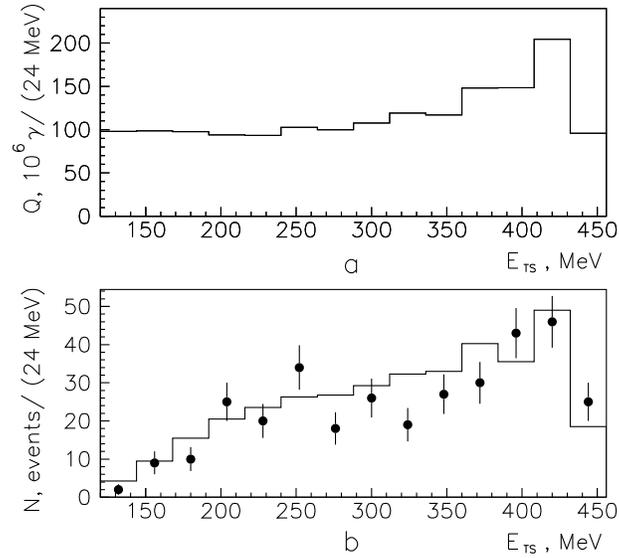}
\caption{(a) The photon energy spectrum  measured in the tagging
system (TS). (b) The number of reconstructed photon splitting
events as a function of the tagged photon energy $E_{TS}$. In plot
(b) black circles present the  experimental results, histogram is
the result of Monte-Carlo simulation based on the exact in
$Z\alpha $ photon splitting cross section.}
\label{Experiment:Fig:expspec}
\end{figure}

\begin{figure}[htb]
\centering
\includegraphics*[width=0.8\textwidth]{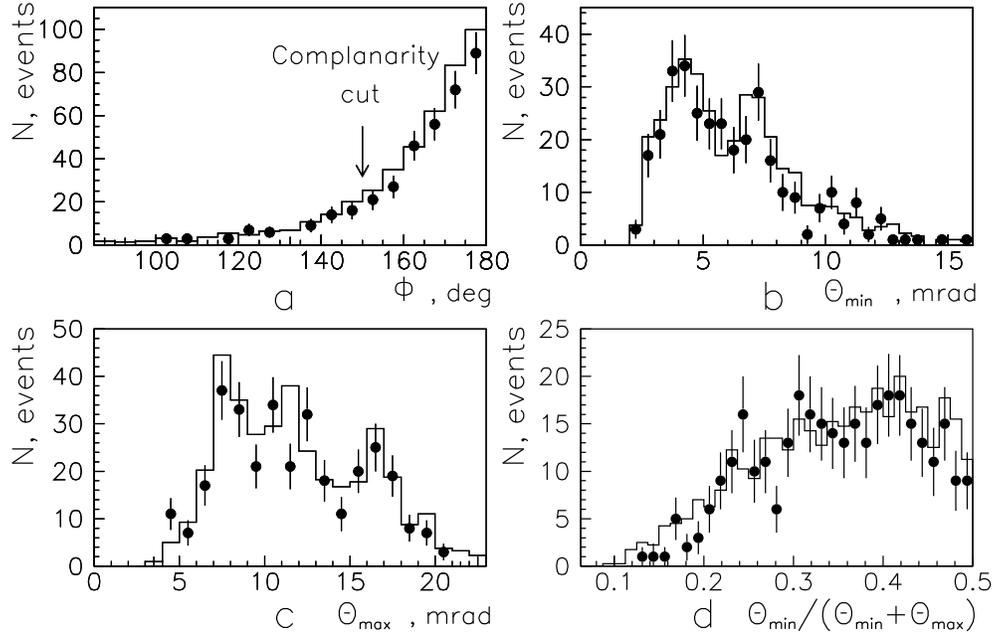}
\caption{The number  of the selected photon splitting events as a
function of
\newline
the azimuth angle between momenta of the outgoing photons (a);
\newline
 the polar angle $\theta_{min}=\mbox{min}\{\theta_2,\theta_3\}$
(b);
\newline
 the polar angle $\theta_{max}=\mbox{max}\{\theta_2,\theta_3\}$
(c);
\newline
 the variable $\tilde
x={\theta_{min}}/({\theta_{min}+\theta_{max}})$ (d).
\newline
 In figures
(b), (c), and (d) only events satisfying the complanarity cut
$\varphi \ge 150^\circ $ (see plot (a)) are included. Black
circles present the  experimental results, histograms are the
results of Monte-Carlo simulation based on the exact in $Z\alpha $
photon splitting cross section.} \label{Experiment:Fig:pspic98}
\end{figure}

The experimental results are presented in Tab.~\ref{tab:res} and
in Figs.~\ref{Experiment:Fig:expspec},
\ref{Experiment:Fig:pspic98} together with the results of
Monte-Carlo simulation based on the exact photon splitting cross
section. The energy spectrum of the initial photons measured in
the tagging system is shown in
Fig.\ref{Experiment:Fig:expspec}(a). Tab.~\ref{tab:res} and
Fig.~\ref{Experiment:Fig:pspic98} present the data summed up over
the initial photon spectrum. The errors shown in the Table are
statistical ones. The systematic error is determined by the
accuracy  of the measurement of the number of initial photons  and
by the uncertainty in the reconstruction efficiency of photon
splitting events. The estimation of these systematic errors gives
2~\% and 5~\%, respectively.

\begin{table}[h]
\centering  \caption{The number of reconstructed events meeting
the selection criteria. Here Q is the number of incoming photons.
The quantity $N_{\varphi<150^\circ }$ is the number of events with
the complanarity angle~${\varphi<150^\circ }$ (see
Fig.~\ref{Experiment:Fig:pspic98}), $N_{\varphi>150^\circ }$ is
the number of events with ${\varphi>150^\circ }$. The quantities
$N_{\varphi<150^\circ}$ and $N_{\varphi>150^\circ}$ are normalized
to the experimental statistics collected with the target. MC means
Monte-Carlo simulation.} \vspace{0.5cm}
\begin{tabular}{|c|c|c|c|c|}
\hline {\bf DATA}&{\bf TARGET}&${\bf Q,\,{10^{9}}}$
&$\bf N_{\varphi>150^\circ }$ & $\bf N_{\varphi<150^\circ }$ \\
\hline Experiment  &$\rm Bi_{4} Ge_{3} O_{12}$ &1.63 & 336$\pm$18
& 82$\pm$9 \\
Experiment   &no target               &0.37 &10$\pm$3
&10$\pm$3 \\
MC photon splitting &$\rm Bi_{4} Ge_{3} O_{12}$ &6.52 &364$\pm$10
& 72$\pm$5  \\
MC Delbr\"uck scattering &$\rm Bi_{4} Ge_{3} O_{12}$ &1.63
&2$\pm$1
& 16$\pm$4  \\
 MC other backgrounds  &$\rm Bi_{4} Ge_{3} O_{12}$ &1.63
&0
& 16$\pm$4  \\
\hline
\end{tabular}
\label{tab:res}
\end{table}

As seen from the Table \ref{tab:res} and
Fig.~\ref{Experiment:Fig:pspic98}a, the main part of the photon
splitting events has, in agreement with the theory, a complanarity
angle $\varphi$ (the azimuth angle between final photon momenta)
close to $180^\circ$. The choice of the interval
$\varphi>150^\circ$ allows us to improve the signal-to-background
ratio (see, e.g., the last two rows of the Table \ref{tab:res}).
Just this $\varphi$-interval was used to plot the distributions
over polar angles in Fig.~\ref{Experiment:Fig:pspic98} and the
dependence of the number of reconstructed photon splitting events
on the initial photon energy $E_{TS}$ in Fig.~\ref
{Experiment:Fig:expspec}(b). Note that for most of the events in
this $\varphi$-interval, the variable
 $\tilde x={\theta_{min}}/({\theta_{min}+\theta_{max}})$ is
 approximately equal to the ratio
 $\mbox{min}(\omega_2,\,\omega_3)/\omega_1$ since the main
 contribution to the cross section comes from the region
 $|{\bf k}_{2\perp}+{\bf k}_{3\perp}|\ll {k}_{2\perp},\,{k}_{3\perp}$,
  i.e. $\varphi\approx 180^\circ$ and
  $\omega_2\theta_2\approx\omega_3\theta_3$.

The results presented in the Table~\ref{tab:res} and in
Figs.~\ref{Experiment:Fig:expspec}(b),
\ref{Experiment:Fig:pspic98} show a good agreement between the
theory and the experiment. More precisely, the total number of
reconstructed  events in the experiment (see Table~\ref{tab:res})
differs from the result of MC simulation by 1.5 standard
deviations.

\begin{figure}[tb]
\centering
\includegraphics*[width=0.7\textwidth]{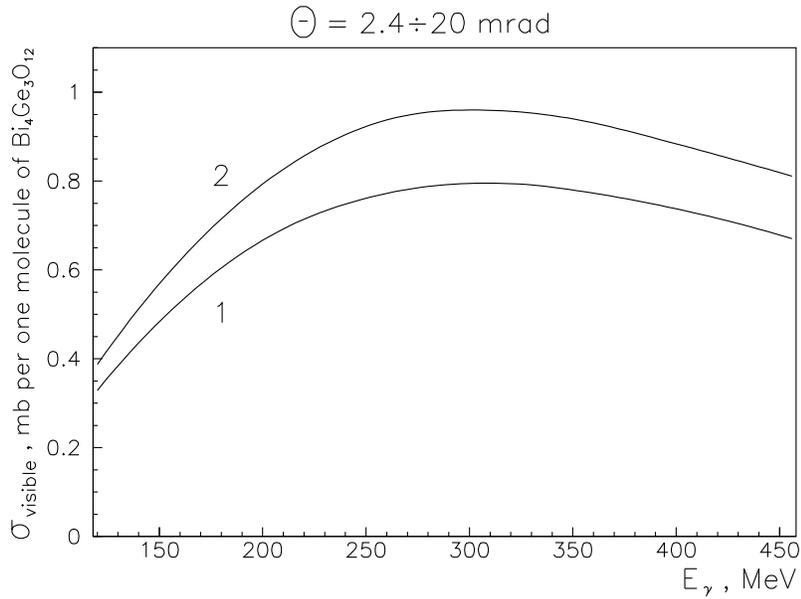}
\caption{The visible photon splitting cross section calculated
exactly in $Z\alpha$ (1) and in the Born approximation(2) as a
function of the initial photon energy.}
\label{Experiment:Fig:BornCoulomb}
\end{figure}

In order to demonstrate the role of the Coulomb corrections under
the experimental conditions, we show in
Fig.~\ref{Experiment:Fig:BornCoulomb} the visible  cross sections
calculated exactly in $Z\alpha$  and in the Born approximation as
a function of the initial photon energy. For all energies
considered, the Born result exceeds the exact one by  20~\%
approximately. Therefore, the use of the Born cross section for MC
simulation would lead to the disagreement between the theory and
the experiment of 3.5 standard deviations. In other words, the
experimental results are significantly closer to predictions of
the exact theory than to those obtained in the Born approximation.

\section{Conclusion}

The results obtained confirm the existence of photon splitting
phenomenon. They also  make possible  the quantitative comparison
with the theoretical predictions. Moreover, the attained
experimental accuracy allows one to distinguish between the
theoretical predictions obtained with or without accounting for
the Coulomb corrections. It turns out that the Coulomb corrections
essentially improve the agreement between the theory and the
experiment. We conclude that the experiment and the theory are
consistent within the achieved experimental accuracy.

\section*{ Acknowledgments}
We are grateful to the staff of the VEPP-4M accelerator complex
for reliable work during the data taking. We thank A.N.~Skrinsky
and V.A.~Sidorov for support of this experiment. Partial support
by the RFBR (Grant 01-02-16926) is also gratefully acknowledged.

\newpage

\end{document}